\documentclass{iau} 
\usepackage{graphicx}

\title[Galaxy Zoo] 
{Twelve Years of Galaxy Zoo}

\author[Karen L. Masters]   
{Karen L. Masters$^{1}$
 \and Galaxy Zoo Team}

\affiliation{$^1$Haverford College, Department of Physics and Astronomy, 370 Lancaster Avenue, Haverford, PA 19041, USA \\ email: {\tt klmasters@haverford.edu}}

\pubyear{2019}
\volume{353}  
\jname{Galactic Dynamics in the Era of Large Surveys}
\editors{M. Valluri, \& J. A. Sellwood  }
\begin{document}

\maketitle

\begin{abstract}
The {\it Galaxy Zoo} project has provided quantitative visual morphologies for over a million galaxies, and has been part of a reinvigoration of interest in the morphologies of galaxies and what they reveal about galaxy evolution. Morphological information collected by GZ has shown itself to be a powerful tool for studying galaxy evolution, and GZ continues to collect classifications - currently serving imaging from DECaLS in its main site, and running a variety of related projects hosted by the {\it Zooniverse}; the citizen science platform which came out of the early success of GZ. I highlight some of the results from the last twelve years, with a particular emphasis on linking morphology and dynamics, look forward to future projects in the GZ family, and provide a quick start guide for how you can easily make use of citizen science techniques to analysis your own large and complex data sets. 

\end{abstract}

\firstsection 
\section{Introduction}

It has been more than twelve years since the launch of {\it Galaxy Zoo} (on 11th July 2007)\footnote{\tt www.galaxyzoo.org}, and the project has been continuously active during that period, inviting volunteers to help provide galaxy classifications, using images from a variety of different sources (see Table \ref{tab1}); and with the science team making use of the aggregated and reduced classifications to investigate a wide range of topics in extragalactic astrophysics. 

 This invited talk for the IAU Symposium, ``Galactic Dynamics in the Era of Large Surveys" was to cover ``Key Results from {\it Galaxy Zoo}". As well as providing a brief overview of twelve years of the project, I will focus on results (given the topic of the conference) most linked to disc galaxy dynamics and the structures this creates, and also look to the future plans for {\it Galaxy Zoo} and the {\it Galaxy Zoo} method. 
 
\section{The Galaxy Zoo Method}

{\it Galaxy Zoo} was inspired by the 1 million galaxy sample of the original Sloan Digital Sky Survey (SDSS; namely the Main Galaxy Sample of \cite{Strauss2002}, which was fully released in Data Release 7, \cite{DR7}). This was the original data set, and the original {\it Galaxy Zoo} (\cite{Lintott2008}) asked for simple classification into spiral or not; further asking spirals to be identified as ``S" or ``Z" shaped (to test claims of a preferred sense of rotation; \cite{Land2008}), or edge-on/can't tell, and had categories for merging systems (and artefacts). This sample was classified with a median of 40 people per image in about eighteen months of operations, and the aggregated classifications were released in \cite{Lintott2011}. 

\begin{figure}[b]
\begin{center}
\vspace*{-2.0 cm}
 \includegraphics[width=4.0in,angle=-90]{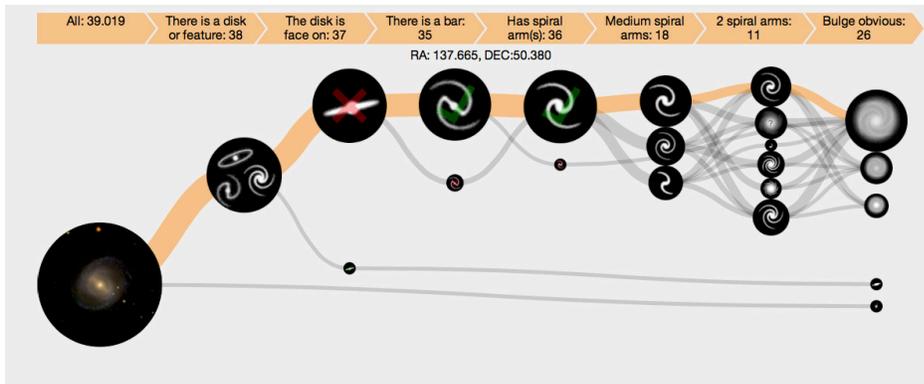} 
 \vspace*{-1.5 cm}
 \caption{An example classification from {\it Galaxy Zoo} 2 in the format which is available at {\tt www.visualise.galaxyzoo.org}. This galaxy is NGC 2771, which has a classification of (R')SB(r)ab in the RC3 (\cite{RC3}). The numbers in the orange bar at top represent the number of users (after a weighting scheme is applied which removes users who always provide highly inconsistent answers) who responded positively to the question indicated.}
   \label{fig:classification}
\end{center}
\end{figure}

 Following the success of the first phase, {\it Galaxy Zoo 2}, (GZ2) was launched, asking for more detailed classifications for the brightest ($r<16$) 250,000 of the SDSS sample. Figure \ref{fig:classification} shows an example set of responses for galaxy NGC 2771 from GZ2\footnote{See {\tt www.visualise.galaxyzoo.org} for more examples}. The GZ2 classifications were reduced and released in \cite{Willett2013}; an improved treatment of redshift bias on the spiral arm questions is presented in \cite{Hart2016}, and these improved classifications are now also available.   

Following the completion of classifications on the original SDSS imaging, {\it Galaxy Zoo} relaunched with all public images from optical Hubble Space Telescope surveys (COSMOS, GOODS, etc; these classifications were published in \cite{Willett2017}), as well as the images from the CANDELS survey (classifications available in \cite{Simmons2017}). Other survey images for which {\it Galaxy Zoo} have obtained classifications include UKIDSS and KiDS; these are not yet published. We have also shown images generated by the Illustris simulation (see \cite{Dickinson2018}), with the goal of testing how well such large scale simulations do at reproducing the population demographics of galaxy morphology. Currently live on the site are images from the Dark Energy Camera Legacy Survey (DECaLS; also see Section \ref{future}). 
 
\begin{table}
  \begin{center}
  \caption{Summary of Data Sets and Projects in the {\it Galaxy Zoo} Project}
  \label{tab1}
 {\scriptsize
  \begin{tabular}{|l|c|c|c|c|}\hline 
{\bf Project} & {\bf Size} & {\bf Imaging Source/} & {\bf Data Access} & {\bf Reference} \\ 
   & (approx) & {\bf Classification Type} & & 
\\ \hline
Original {\it Galaxy Zoo}  &1 million & SDSS Main Galaxy Sample  & Public$^{1,2}$ & \cite{Lintott2011} \\
& & (spiral/early-type/mergers only) \\
\hline
{\it Galaxy Zoo} 2 & 250,000 & SDSS Main Galaxy Sample $r<17$. & Public$^{1,2}$ & \cite{Willett2013} \\
& & Full classification tree. \\
& & & & \cite{Hart2016} \\ \hline
{\it Galaxy Zoo}: Hubble & 120,000 & Public Optical HST Surveys & Public$^{1}$ & \cite{Willett2017} \\
& & (COSMOS, GOODS etc) & \\ \hline
{\it Galaxy Zoo}: CANDELS & 48,000 & CANDELS Survey Galaxies & Public$^{1}$ & \cite{Simmons2017} \\ \hline 
{\it Galaxy Zoo}: MaNGA & 30,000 & MaNGA Target Sample & Public$^{1,2}$ & \cite{DR15} \\
& & GZ2 style & &  (SDSS DR15)  \\ \hline
{\it Galaxy Zoo} 4  &  & SDSS Extra Imaging (DR12) $r<17$ & Pending & \\
 {\it Galaxy Zoo}: UKIDSS &  & UKIDSS &  Pending & \\
 {\it Galaxy Zoo}: Illustris & &  Illustris Simulated Galaxies &  Pending & \cite{Dickinson2018} \\
 {\it Galaxy Zoo}: KiDS & & GAMA KiDS & Pending &  \cite{Holwerda2019} \\ \hline
  \end{tabular}
  }
 \end{center}
\vspace{1mm}
 \scriptsize{
{\it Notes:}\\
$^1$Public data is available at {\tt www.data.galaxyzoo.org} \\
$^2$Also available via SDSS Casjobs at {\tt www.skyserver.org}}
\end{table}

\section{Summary of Published Results}

The {\it Galaxy Zoo} science team have published over 60 papers in the last 12 years which make use of the classifications from different phases of GZ\footnote{A complete listing of team papers can be found at \\ {\tt www.zooniverse.org/about/publications}.}  At the time of writing there are twelve papers from the team with more than 100 citations each. 

 Attempting to categorize publications into types, GZ team papers can be broken down into: 
 \begin{itemize}
 \item {\it Data papers} which describe the techniques to go from clicks to classifications (including user weighting, and accounting for image quality, and changes in how galaxies appear in images as a function of redshift). These are: \cite{Lintott2011,Willett2013,Willett2017,Simmons2017}. In these papers you can also find comparisons of {\it Galaxy Zoo} morphologies with other published classifications and morphological proxies (e.g. structural parameters). 
 \item Papers which look at {\it broad statistical properties of the morphologies of galaxies}. These have often focused on the comparison of colour and morphology (see Section \ref{colour}). Examples include: \cite{Bamford2009,Skibba2009,Schawinski2009,Masters2010r,Schawinski2014,Smethurst2015,Hart2016,Masters2019}.
 \item Papers about {\it interesting/odd objects (or classes of objects) revealed by {\it Galaxy Zoo}}. Examples include the green peas (compact ionized galaxies, see \cite{Cardamone2009}) and Hanny's Voorwerp (a light echo from a recently quiescent accreting black hole; see \cite{Lintott2009} and recent work by \cite{Keel2019}). 
 \item Papers which investigate the {\it morphology of specific types of galaxy, or the properties of galaxies with specific morphological features}. Examples include work on the host galaxies of Active Galactic Nuclei (AGN): \cite{Schawinski2010,Simmons2013,Galloway2015,Smethurst2016}; mergers: \cite{Darg2010a,Darg2010b} or many papers on the properties of disc galaxies with bars: \cite{Masters2011,Masters2012,Cheung2013,Melvin2014,Simmons2014,Kruk2018}. Section \ref{bars} will discuss our work on bars in more detail. 
 \end{itemize}

\begin{figure}[b]
\vspace*{-1.5 cm}
\begin{center}
 \includegraphics[width=3.6in,angle=-90]{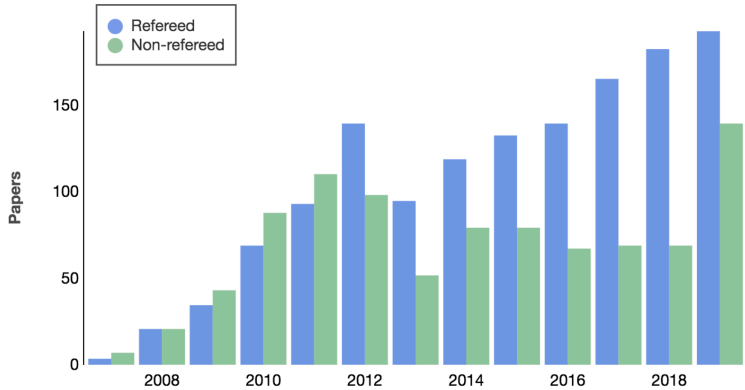} 
\vspace*{-1.5 cm}
 \caption{Count of papers (refereed in blue, non-refereed in green) mentioning the term "{\it Galaxy Zoo}" over the last twelve years, since {\it Galaxy Zoo} launched in 2007. Credit: NASA/SAO Astrophysics Data System; ADS}
   \label{fig:count}
\end{center}
\end{figure}

In addition to team papers, the public release of classifications has led to many more results making use of GZ classifications. Overall, at the time of writing\footnote{Based on a literature search of the phrase "{\it Galaxy Zoo}"} the classifications have been used or mentioned in over 1000 papers in the last 12 years. Figure \ref{fig:count} shows how this has grown since 2009 (note that 2019 data is not yet complete). 

\section{Colour and Morphology} \label{colour}

One of the earliest contributions to come out of GZ classifications, was putting on a firm statistical footing the correlation between galaxy colour and morphology. While it had been known for a long time (e.g. as commented by \cite{Zwicky1955}) that spiral galaxies were bluer than elliptical galaxies, it was the advent of large galaxy surveys (most notably SDSS) that let to the common perception that colour and morphology were entirely equivalent. Although this result was based on fairly small samples of galaxies with morphological classifications, it remains fairly common in the astronomical literature today. 

\begin{figure}[b]
\vspace*{-6.0 cm}
\begin{center}
 \includegraphics[width=5in,angle=-0]{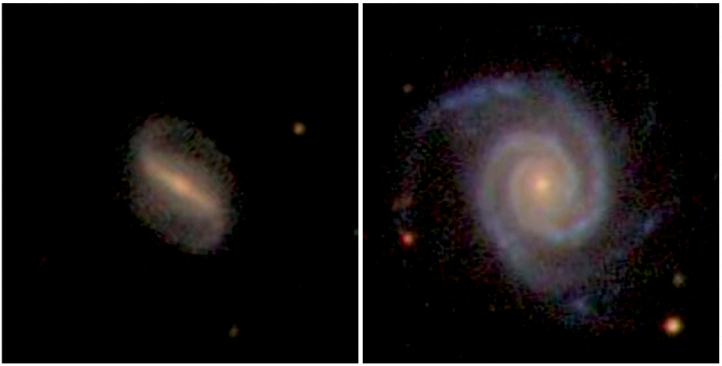} 
 \vspace*{-5.0 cm}
 \caption{Example images of red (left) and blue (right) spiral galaxies identified in {\it Galaxy Zoo}. These also demonstrate the barred/not barred selection from GZ2.}
   \label{fig:morphology}
\end{center}
\end{figure}

 In a pair of early papers, GZ classifications were used to look at the correlation between colour in morphology in a larger sample than ever before (\cite{Skibba2009,Bamford2009}). These papers revealed the existence of significant numbers of red spirals and blue ellipticals which were investigated in \cite{Masters2010r} and \cite{Schawinski2010} respectively. Red spirals in particular were found to be quite common, with more than half of the most massive spirals being optically red (even when removing highly inclined spirals, and very early-type spirals with large bulges). Example images of red and blue spirals are shown in Figure \ref{fig:morphology}. This demonstrated that it is important to recognise that selections based on colour cannot be interpreted as cleanly dividing the population into the two broad morphological types. 
 
The imperfect correlation between colour and morphology makes sense physically. The morphology of a galaxy (to ``first order", ignoring things like dust) provides information on the orbits of stars within it. As such, is should be obvious that important clues to the formation history of galaxies is revealed by their morphologies, and this information is complimentary, but not identical to, their star formation history and chemical composition as revealed by photometry (i.e. colour) and spectra. 

\section{Bars and Spiral Arms} \label{bars}

The more detailed classifications of GZ2 enabled the study of features in disc galaxies tied intimately to their dynamics, such as bars, and spirals. 

\subsection{Bars}
The GZ team have also published a wide range of papers investigating bars in galaxies. Early results on bars revealed something unexpected at the time, that they were more common in redder spirals (\cite{Masters2011}). Work to compare what kinds of bars are identified by GZ classifications (with a threshold of $p_{\rm bar} > 0.5$) demonstrated they were strong bars (SB in the traditional notation, but not SAB); while weaker bars can be identified by lower vote fractions this is at the risk of including larger numbers of erroneous classifications; see e.g. \cite{Willett2013}). Example images of a red barred spiral and a blue unbarred spiral are shown in Figure \ref{fig:morphology}. Further work has confirmed a picture of strong bars being preferentially associated with massive, red, gas poor (\cite{Masters2012}) and low star forming (\cite{Cheung2013}) discs, and in addition showing evidence for secular growth of pseudo-bulges (\cite{Cheung2015}). Classifications on the public HST images, and CANDELS allowed investigation of bar fractions to be extended to higher redshift revealing how they become more common in the later Universe (\cite{Melvin2014,Simmons2014}). The clustering of GZ2 barred galaxies was investigated in \cite{Skibba2012}, while  \cite{Galloway2015} studied the impact galactic scale bars had on AGN properties, and \cite{Kruk2017,Kruk2018} revealed the presence of offset bars, and provided three component (bar, bulge, disc) photometric fits for a large sample from GZ2.

\subsection{Spiral Arms}
Spiral arms (the number and winding angle) are also described in GZ2 classifications. The properties of spiral arms in the population has been investigated in a series of recent papers. \cite{Hart2016,Hart2017a} provided an updated method to correct classifications for redshift bias, and showed that many arm spirals are associated with recent star-formation, and also that the distribution of SF differs between two-armed (``grand design") and many armed (``flocculent") spirals. The winding properties identified in GZ2 were then used to investigate spiral formation mechanisms; interestingly we find that the classic static density wave does not match GZ2 classifications on a population scale -- they reveal a lack of strong correlation between bulge size and spiral arm pitch angle predicted by those models (\cite{Hart2017b,Hart2018,Masters2019}) . Further work on this topic may reveal many interesting further constraints on spiral arm formation mechanisms.

\section{Spin Off Projects}

The GZ technique of crowdsourcing the analysis of galaxy images can also be applied to more than just answering simple questions. New capabilities in {\it The Zooniverse Project Builder} (see Section 8) make this very straightforward now, but even before this existed there were efforts to make spin-off projects linked to GZ by science team members. Spin-off {\it Galaxy Zoo} projects run by the science team in part, or whole include: 
\begin{itemize}
\item {\it Galaxy Zoo Mergers} - a custom built project in which volunteers were asked to run models of galaxy mergers and find the best match to images. (\cite{Holincheck2016})
\item {\it Galaxy Zoo Bar Lengths} - this project ran in a Google interface to {\it Galaxy Zoo} and measured bar lengths on SDSS images. For details see \cite{Hoyle2011}. A second {\it Galaxy Zoo Bar Lengths} was run more recently\footnote{https://www.zooniverse.org/projects/vrooje/galaxy-zoo-bar-lengths} to do expand on this work using Hubble images; results will be presented in Hutchinson-Smith et al. in prep. 
\item {\it Galaxy Zoo: 3D} - this spinoff uses {\it Project Builder} to display images of galaxies which are part of the SDSS-IV project, Mapping Nearby Galaxies at Apache Point Observatory (MaNGA; \cite{Bundy2015}), and is asking users to draw the locations of bars, spiral arms and foreground stars (as well as mark the centers of the galaxies)\footnote{https://www.zooniverse.org/projects/klmasters/galaxy-zoo-3d}. The resulting masks can be used with MaNGA data to extract spectra from the respective features (e.g. as used by \cite{Peterken2019,FraserMcKelvie2019}). The project will be described in full in an upcoming Masters et al. in prep. 
\item {\it Galaxy Builder} - Another challenge for extragalactic astrophysicists is decomposing galaxy images into multiple components. While large databases exist of simple automated fits, significant human interaction is generally still required to ensure physical results. An attempt to involve crowdsourcing in this process is Galaxy Builder\footnote{\tt https://www.zooniverse.org/projects/tingard/galaxy-builder}, which makes use of crowdsourced galaxy models to select the number of needed componnts (from disc, bulge, bar, and any number of spirals) and then initiate a final automated fit. This project will be described in full in an upcoming Lingard et al. in prep.. and is a custom project built inside the {\it Project Builder} framework. 
\item {\it Clump Scout} - very recently launched\footnote{\tt https://www.zooniverse.org/projects/hughdickinson/galaxy-zoo-clump-scout} this project is asking volunteers to help identify star formation clumps in galaxies. 
\end{itemize}

\section{The Future of Galaxy Zoo - Machines and Humans Working Together}\label{future}

For {\it Galaxy Zoo} itself, the next generation of sky surveys, moving to both deeper and higher resolution will challenge the volume of galaxies which are classifiable by the crowd. Luckily machine learning algorithms are also advancing. An initial study by \cite{Beck2018}  demonstrated the use of a random forest algorithm to significantly speed up the first (binary) step on the question tree. Today, the main GZ site is running a new ``Enhanced Workflow", which uses a Bayesian model to prioritze the galaxies shown to volunteers. A Bayesian Convolutional Neural Network trains on the crowdsources classifcations at the end of each week, and quickly becomes expert at classifying the simplest galaxies, while identifying more complex galaxy which need to be shown to the human volunteers for further classification. For more details on this see \cite{Walmsley2019}. 

\section{How to Build Your Own Zooniverse Project}

The early success of {\it Galaxy Zoo} as a public engagement project (at its peak obtaining 70,000 classifications an hour; and noted by {\it The Guardian} in 2012\footnote{\tt http://www.guardian.co.uk/science/2012/mar/18/galaxy-zoo-crowdsourcing-citizen-scientists} as providing ``no surer way to engage the public than to involve people in the research itself") inspired the creation of {\it The Zooniverse}\footnote{{\tt www.zooniverse.org}}, a platform which hosts similar citizen science projects across any research topic. {\it The Zooniverse} launched in 2010 to make a common community of citizen scientists, and now has over 1.7 million account holders. In the early days each project was custom built by the development team, which limited the rate at which projects could launch. A significant innovation came with the launch of {\it Project Builder}\footnote{\tt https://www.zooniverse.org/lab} in 2015. This facility allows researcher to build their own projects, which can be submitted to  {\it The Zooniverse} for peer review and beta testing if they wish to be promoted to the millions of account holders, or simply run independently. A simple project based on a few images can be built in a matter of minutes, entirely in the browser, and facilties exist to provide several different types of common crowdsourcing activities (e.g. question trees, drawing, annotation). A total of just over 200 projects have been formally launched by {\it The Zooniverse}, out of more than 9000 which have been built on {\it Project Builder} (300 of which have independently collected more than 100 classifications). 

{\it Zooniverse Project Builder} projects don't have to be public facing. One example of use within astronomy was for the visual inspection of spectra obtained for the MaNGA Spectral Library (\cite{Yan2019}), where 28 members of the MaNGA science team helped to visually inspect the quality of 10,797 stellar spectra.

\section{Summary and Conclusions}

In this short paper, based on an invited review talk given at the ``Galactic Dynamics in the Era of Large Surveys" conference in Shanghai, China in July 2019 I give a whistle-stop tour through the first twelve years of the {\it Galaxy Zoo} project, hi-lighting in particular results related to the imperfect correlation between galaxy colour and morphology, as well as results related to the disc galaxy dynamical structures: galactic bars and spirals. Almost a teenager now, {\it Galaxy Zoo} is a mature technique, which has had a significant impact on the field of extragalactic astrophysics. New innovations in both crowdsourcing, and the combination of humans and computers working together ensure that the project is still relevant and productive today, has a long future to come.


\end{document}